\documentclass[aps,pra,twocolumn,superscriptaddress,nofootinbib]{revtex4-2}

\usepackage{placeins}
\usepackage{graphicx}
\usepackage[utf8x]{inputenc}
\usepackage[T1]{fontenc}
\usepackage[nointegrals]{wasysym}
\usepackage{xcolor}
\usepackage{amsmath}
\usepackage{bbold} 
\usepackage{mathtools}
\usepackage[normalem]{ulem}

\begin{document}

\title{Enhancing collective entanglement witnesses through correlation with state purity}

\author{Kateřina Jiráková} 
\affiliation{Institute of Physics of the Academy of Sciences of the Czech Republic, Joint Laboratory of Optics of Palacký University and Institute of Physics AS CR, 17. listopadu 50a, 772 07 Olomouc, Czech Republic}

\author{Antonín Černoch} 
\affiliation{Institute of Physics of the Academy of Sciences of the Czech Republic, Joint Laboratory of Optics of Palacký University and Institute of Physics AS CR, 17. listopadu 50a, 772 07 Olomouc, Czech Republic}

\author{Artur Barasiński} 
\affiliation{Institute of Theoretical Physics, University of Wroclaw, Plac Maxa Borna 9, 50-204 Wroclaw, Poland}

\author{Karel Lemr} \email{k.lemr@upol.cz}
\affiliation{Palacký University in Olomouc, Faculty of Science, Joint Laboratory of Optics of Palacký University and Institute of Physics AS CR, 17. listopadu 12, 771 46 Olomouc, Czech Republic}


\begin{abstract}

This paper analyzes the adverse impact of white noise on collective quantum measurements and argues that such noise poses a significant obstacle to the otherwise straightforward deployment of collective measurements in quantum communications. The paper then suggests addressing this issue by correlating the outcomes of these measurements with quantum state purity. To test the concept, a support vector machine is employed to boost the performance of several collective entanglement witnesses by incorporating state purity into the classification task of distinguishing entangled states from separable ones. Furthermore, the application of machine learning allows to optimize selectivity of entanglement detection given a target value of sensitivity. A response operating characteristic curve is reconstructed based on this optimization and the area under curve calculated to assess the efficacy of the proposed model.

\end{abstract}

\date{\today}

\maketitle

\section{Introduction}
\begin{figure}[t]
\centering
\includegraphics[scale=1]{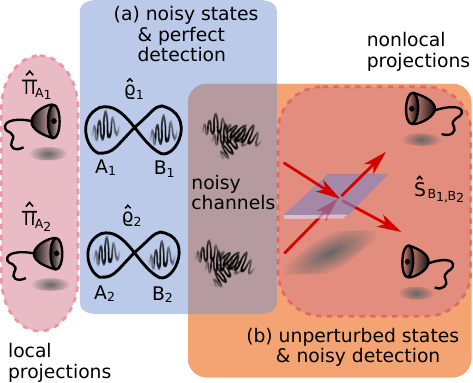}
\caption{Conceptual scheme of collective measurement with $n = 1$ nonlocal projections. Imperfections in the setup can be modelled by insertion of two noisy channels. The effect of these channels can be described by two equivalent strategies: (a) the two states $\hat{\varrho}_1$ and $\hat{\varrho}_2$ become noisy while the nonlocal measurement remains perfect or (b) perfect unperturbed states are subject to imperfect nonlocal projection.}
\label{fig_conc_scheme}
\end{figure}

Collective quantum measurements, which are measurements performed simultaneously on multiple copies of the investigated state, are an invaluable tool in quantum state analysis. Since the pioneering experimental demonstration in 2005~\cite{Bovino2005}, collective measurements have been instrumental in practical implementation of a number of entanglement witnesses~\cite{Rudnicki_2011,Rudnicki_2012,Bartkiewicz.PhysRevA.95.2017,Bartkiewicz_2017,Bartkiewicz.PhysRevA.97.2018,Travnicek_2018,Roik_2021,ROIK_2022} and to infer quantum state purity or fidelity~\cite{Travnicek_2019,Bartkiewicz_PhysRevA.99}. The key advantage of these measurements lies in the projection of subsystems from different copies of the examined state onto a common entangled state (see nonlocal projection in conceptual diagram in Figure~\ref{fig_conc_scheme}). Thanks to these nonlocal projections, collective measurements outperform ordinary single-copy measurements in terms of sensitivity (e.g. the volume of detected entangled states~\cite{Bartkiewicz.PhysRevA.95.2017}) or in the efficiency in achieving set tasks with fewer measurements~\cite{Roik_2021,ROIK_2022}. Notably, the required number of collective measurements does not grow as prohibitively fast with the size of the Hilbert space as is the case with standard quantum state tomography.

Application-wise, collective measurements hold particular appeal in conjunction with entanglement distribution in quantum networks. Their layout (visualized in Figure~\ref{fig_conc_scheme}) aligns with the topology of entanglement swapping making collective measurements straightforwardly deployable in quantum repeaters~\cite{bib:Briegel:repeater}, relays~\cite{bib:Jacobs:relay} and in teleportation-based quantum communications networks~\cite{Yonezawa2004,PhysRevLett.122.170501}. For instance, research has demonstrated applicability of collective measurements for a rapid and practical diagnostics of the entanglement swapping protocol~\cite{Travnicek_2020}.
Moreover, direct measurement of Hilbert-Schmidt distance through collective measurements has been used to speed-up K-means classification algorithm~\cite{Travnicek_2019}.

As explained above, the power of collective measurements lies in the nonlocal projections. However, these projections are notably susceptible to white noise, and consequently, utilization of collective measurements in real-world quantum communications necessitates an assessment of their performance in the presence of that noise. In Sec.~\ref{sec:noise} of this paper, we analyze propagation
of the investigated state through noisy channel and document the adverse effect of the channel on collective measurements as the noise accumulates polynomially fast with the power corresponding to the number of copies projected onto a shared entangled state. Hence even a rather weakly perturbing realistic quantum channel can cause the projections to become local (no longer projecting on an entangled state) and, thus, the entire collective measurement looses its power. This effect is exemplified on the incapability of collectibility (a collective entanglement witness) to correctly detect entangled Werner state with a noise level above $p = 1-\tfrac{\sqrt{3}}{2} \approx 0.13$~\cite{Rudnicki_2012}.

Simultaneously, white noise is commonly associated with a reduction in the purity of transmitted quantum states. In this paper, we analyze the correlation between collective entanglement witnesses and purity of the states aiming to mitigate the influence of white noise on collective measurements. Our findings reveal that this way one can considerably enhance sensitivity of collective entanglement witnesses (true positive rate~\cite{Sheskin_2003,Dekking_2005} or TPR) at a relatively small cost in decreased selectivity (false positive rate or FPR). A support vector machine is employed to find the optimal classification boundary between entangled and separable states, maximizing selectivity for a given value of sensitivity boost (see Sec.~\ref{sec:svm}). Furthermore, targeting several sensitivity values and identifying the corresponding selectivity allows us to reconstruct the entire receiver operating characteristic (ROC) curve, evaluating the method's effectiveness in terms of the area under this curve (AUC). Note that reconstructing the ROC curve would not be possible without the use of machine learning as analytical entanglement witnesses assume perfect selectivity (FPR$=0$). However, such assumption is unrealistic due to the inevitable technical imperfections in practical conditions, at least such as detection shot noise~\cite{Perina1991}. Our analysis of the ROCs contributes to the assessment of practicality of collective entanglement witnesses for near-future quantum communications networks.

\section{Vulnerability of collective measurements to white noise}
\label{sec:noise}
In this section, we demonstrate that collective measurements are vulnerable to the presence of noise, in particular white noise. 
Let us consider a simple scenario illustrated in the Figure~\ref{fig_conc_scheme} and assume that two copies of a Bell state $\hat{\varrho}_{1} = \hat{\varrho}_{2} = \hat{\text{P}}^+=|\phi^{+} \rangle \langle \phi^{+}|$ ($|\phi^{+}\rangle=\frac{1}{\sqrt{2}} \sum_{i=0}^1 |ii\rangle$) are subject to the collective measurement. 
Imperfections in the procedure are modelled by two channels introducing the noise with probability $p$. There are two equivalent strategies that can be used to incorporate the effect of noise on the outcome of collective measurements. In strategy (a) we add noise to the investigated states $\hat{\varrho}_{1}$, $\hat{\varrho}_{2}$ that are then tested in a perfect collective measurement. In the second strategy (b), both noisy channels are added to the nonlocal measurement projector $\hat{S}_{\text{B}_{\text{1}}, \text{B}_{\text{2}}}$ turning it into a POVM. In this strategy, perfect Bell states $\hat{\varrho}_{i}$ are measured by means of this POVM. 

To prove this, let us first define the probabilities of collective projections, which in the next section will be used as a central quantity in the analyzed nonclassical witnesses. The probabilities of collective projections are given by 
\begin{equation*}
    C (\hat{\Pi}_{\text{A}_{\text{1}}},\hat{\Pi}_{\text{A}_{\text{2}}}) = \text{Tr}[(\hat{\chi}_{\text{A}_{\text{1}}, \text{B}_{\text{1}}} \otimes \hat{\chi}_{\text{A}_{\text{2}}, \text{B}_{\text{2}}}) (\hat{S}_{\text{B}_{\text{1}}, \text{B}_{\text{2}}}\otimes\hat{\Pi}_{\text{A}_{\text{1}}}\otimes\hat{\Pi}_{\text{A}_{\text{2}}})] 
\end{equation*}
and
\begin{equation*}
     \bar{C} (\hat{\Pi}_{\text{A}_{\text{1}}},\hat{\Pi}_{\text{A}_{\text{2}}}) = \text{Tr}[(\hat{\chi}_{\text{A}_{\text{1}}, \text{B}_{\text{1}}} \otimes \hat{\chi}_{\text{A}_{\text{2}}, \text{B}_{\text{2}}}) (\hat{\mathbb{1}}_4 \otimes \hat{\Pi}_{\text{A}_{\text{1}}}\otimes\hat{\Pi}_{\text{A}_{\text{2}}})],
\end{equation*}
where $\hat{\Pi}_{i}$ stands for local projection, $\hat{S}_{\text{B}_{\text{1}}, \text{B}_{\text{2}}}$ represents the two-qubit collective measurement, and $\hat{\chi}_{\text{A}_{\text{i}}, \text{B}_{\text{i}}}$ denotes the Bell state $\hat{\varrho}_{i}$ already affected by the noise.
Here, indexes $\text{A}_{i}$ and $\text{B}_{i}$ are used to denote the first and the second qubit of analyzed state.

Note that by virtue of the Jamiołkowski isomorphism \cite{JAMIOLKOWSKI1972275}, the noisy channel $\hat{\chi}$ can be written as $(\hat{\mathbb{1}} \otimes \boldsymbol{\Gamma})[\hat{\text{P}}^+]$, with $\boldsymbol{\Gamma}[\bullet]:=\sum_k \Gamma^{(k)} \bullet (\Gamma^{(k)})^{\dagger}$ a completely positive map (see e.g. \cite{Badziagpra62_2000}). Then, with straightforward calculations, one can write
\begin{eqnarray}
\hat{\chi}_{\text{A}_{\text{1}}, \text{B}_{\text{1}}} \otimes \hat{\chi}_{\text{A}_{\text{2}}, \text{B}_{\text{2}}} = \sum_{k_1,k_2} (\hat{\mathbb{1}}_{\text{A}_{\text{1}}}\otimes\Gamma^{(k_1)}_{\text{B}_{\text{1}}}\otimes\hat{\mathbb{1}}_{\text{A}_{\text{2}}}\otimes\Gamma^{(k_2)}_{\text{B}_{\text{2}}})\nonumber\\
\times(\hat{\text{P}}^+_{\text{A}_{\text{1}}, \text{B}_{\text{1}}} \otimes \hat{\text{P}}^+_{\text{A}_{\text{2}}, \text{B}_{\text{2}}})(\hat{\mathbb{1}}_{\text{A}_{\text{1}}}\otimes\Gamma^{(k_1)}_{\text{B}_{\text{1}}}\otimes\hat{\mathbb{1}}_{\text{A}_{\text{2}}}\otimes\Gamma^{(k_2)}_{\text{B}_{\text{2}}})^{\dagger}.
\end{eqnarray}
Using this equation and the cyclic property of the trace function, one gets
\begin{equation*}
    C (\hat{\Pi}_{\text{A}_{\text{1}}},\hat{\Pi}_{\text{A}_{\text{2}}}) = \text{Tr}[(\hat{\text{P}}^+_{\text{A}_{\text{1}}, \text{B}_{\text{1}}} \otimes \hat{\text{P}}^+_{\text{A}_{\text{2}}, \text{B}_{\text{2}}}) (\hat{\Omega}_{\text{B}_{\text{1}}, \text{B}_{\text{2}}}\otimes\hat{\Pi}_{\text{A}_{\text{1}}}\otimes\hat{\Pi}_{\text{A}_{\text{2}}})], \nonumber
\end{equation*}
where the noisy collective measurement $\hat{\Omega}_{\text{B}_{\text{1}}, \text{B}_{\text{2}}}= \sum_{k_1,k_2} (\Gamma^{(k_1)}_{\text{B}_{\text{1}}}\otimes\Gamma^{(k_2)}_{\text{B}_{\text{2}}})^{\dagger} \hat{S}_{\text{B}_{\text{1}}, \text{B}_{\text{2}}} (\Gamma^{(k_1)}_{\text{B}_{\text{1}}}\otimes\Gamma^{(k_2)}_{\text{B}_{\text{2}}})$ is the mathematical representation of the strategy (b). Note that $\hat{\Omega}_{\text{B}_{\text{1}}, \text{B}_{\text{2}}}$ accumulates noise from both channels. 
Similar calculations can be made for $\bar{C} (\hat{\Pi}_{\text{A}_{\text{1}}},\hat{\Pi}_{\text{A}_{\text{2}}})$ showing that this quantity is, however, invariant to noise as $\boldsymbol{\Gamma}$ is trace-preserving (i.e. $\sum_k (\Gamma^{(k)})^{\dagger} \Gamma^{(k)}=\hat{\mathbb{1}}$). Therefore, scenario (a) is equivalent to scenario (b) when $\{\hat{\chi}_{\text{A}_{\text{1}}, \text{B}_{\text{1}}} , \hat{\chi}_{\text{A}_{\text{2}}, \text{B}_{\text{2}}},\hat{S}_{\text{B}_{\text{1}}, \text{B}_{\text{2}}}\} \leftrightarrow \{\hat{\text{P}}^+_{\text{A}_{\text{1}}, \text{B}_{\text{1}}} , \hat{\text{P}}^+_{\text{A}_{\text{2}}, \text{B}_{\text{2}}},\hat{\Omega}_{\text{B}_{\text{1}}, \text{B}_{\text{2}}}\}$. 
It is worth mentioning that the above considerations apply to an arbitrary state $\chi$, not just noisy Bell states.

Let us now consider a particular, yet practical, situation when $\boldsymbol{\Gamma}$ stands for a depolarizing channel \cite{Nielsencambridge_2000} and $\hat{S}_{\text{B}_{\text{1}}, \text{B}_{\text{2}}} = \hat{\text{P}}^+_{\text{B}_{\text{1}}, \text{B}_{\text{2}}}$. In this case, the Bell state is transformed into the Werner state with visibility $v=(1-p)$
\begin{equation}
    \hat{\text{P}}^+ \longrightarrow \chi=\hat{\varrho}_{\text{W}}(p) = (1-p)~\hat{\text{P}}^+ + p\frac{\hat{\mathbb{1}}}{4}.
\end{equation}
On the other hand, in the second strategy (b) one gets 
\begin{eqnarray}
    \hat{S}_{\text{B}_{\text{1}}, \text{B}_{\text{2}}} &=& \hat{\text{P}}^+
    \longrightarrow\\
    \hat{\Omega}_{\text{B}_{\text{1}}, \text{B}_{\text{2}}} = (1-p)^{2}~\hat{\text{P}}^+ &+& (2-p)p\frac{\hat{\mathbb{1}}}{4} \equiv  \hat{\varrho}_{\text{W}}\Big((1-p)^2\Big),\nonumber
    \label{eq:strategieBb}
\end{eqnarray}
i.e. a POVM described by the Werner state with visibility $v = (1-p)^{2}$. 

Consider that the Werner state becomes separable for $v \leq \tfrac{1}{3}$ and that projecting on a separable state removes the advantage of collective measurements as such measurement can be decoupled into a series of local single-state copy measurements. The threshold $v = \tfrac{1}{3}$ is reached if the channel introduces noise with probability $p = 1-\sqrt{\tfrac{1}{3}} \approx 0.42$. 
As a consequence, entanglement witnesses based on collective measurement thus rather quickly lose their detection power when white noise is introduced. Note that for instance one of these witnesses, the collectibility, does not even detect all Bell nonlocal Werner states given by $p < 1-\tfrac{1}{\sqrt{2}} \approx 0.29$~\cite{Jirakova_2021}. Collectibility only detects nonlocal correlations for Werner states with $p < 1- \tfrac{\sqrt{3}}{2} \approx 0.13$~\cite{Rudnicki_2012}.

 
\section{States generation and classification}
\label{sec:svm}
As demonstrated in the previous section, white noise has a negative effect on sensitivity of collective entanglement witnesses. Simultaneously, white noise also causes purity of examined quantum states to decrease. This leads us to propose the idea of restoring the performance of collective witnesses by correlating their values with purity of the examined states. To test this idea, the following three steps were implemented: (i) generation of random two-qubit states with uniform distribution in purity, (ii) estimation of three collective entanglement witnesses and purity of the state, (iii) classification of the states via machine learning.

\textit{States generation:} First, we have generated random two-qubit states following the method described in Ref.~\cite{Roik_2021,Maziero2015} and also explained in detail in the Appendix. This method is based on the preparation of a diagonal matrix with uniformly distributed eigenvalues which is then subjected to a random two-qubit unitary evolution. The appropriate SU(4) unitary matrices are generated according to a procedure described in Ref.~\cite{Marcin_Pozniak_1998}. Although such approach provides a uniform distribution of states with respect to the Haar measure, it does not provide a uniform distribution of state purity.
There are also other methods for random state generation proposed in literature each leading to a different sampling of the states hence a different distribution of their purities \cite{JOHANSSON2013,Qiskit}. Several of these methods have been compared in the Appendix.
To tackle this issue, we have decided to post-select from the generated states a subset of 2 million states with uniform distribution of purity. We consider this subset useful for benchmarking as it has considerably increased the presence of pure and quasi-pure states with respect to all tested vanilla random state generation algorithms. As a result, it favours entanglement detection based solely on the entanglement witnesses \emph{without} correlating them with purity and hence the improvement obtained by incorporating purity into the decision is rather underestimated and certainly not overestimated.

\begin{figure}
\includegraphics[scale=1]{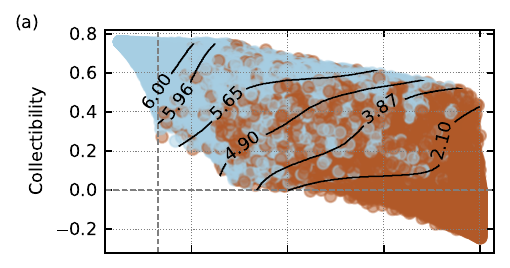}
\includegraphics[scale=1]{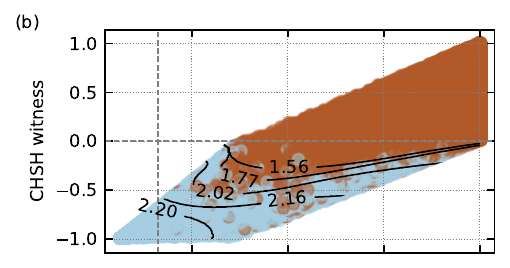}
\includegraphics[scale=1]{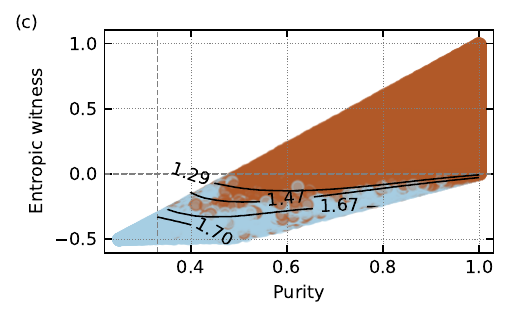}
\caption{\label{fig:contours}Distribution of separable (light blue) and entangled (dark reddish) states in the feature space of the collective witnesses and purity is depicted for the three evaluated witnesses: (a) collectibility, (b) the CHSH witness, and (c) the entropic witness. As a sanity check, the analytical decision threshold at 0 is marked by the horizontal dashed line. Similarly the purity threshold below which all states must be separable at $1/3$~\cite{Rudnicki_2012} is marked by vertical dashed line. Selected support vector machine decision boundaries are plotted by solid black lines labelled by the corresponding improvement factors IF.}\label{fig_distr}
\end{figure}

\textit{Entanglement witnesses and purity:} With the random states generated, we have calculated values of three typical collective entanglement witnesses that can be measured in the entanglement swapping geometry: (a) the collectibility, (b) the CHSH witness, and (c) the entropic witness. 
The collectibility as used in this article is defined as
\begin{eqnarray}
    W &=& \frac{1}{2}\Big[\eta + X^2_0(1-2 X_{00}) + X^2_1(1-2 X_{11}) + \\
    &+& 2 X_0X_1(1-2 X_{01})  -1  \Big]\nonumber
\end{eqnarray}
where
\begin{equation}
\eta = 16 X_{0} X_{1}\sqrt{X_{00} X_{11}} + 4 \cdot\text{max} \{X_{++}, X_{--}\} 
\end{equation}
and
\begin{eqnarray}
    X_{ij} &=& \frac{ C(\hat{\Pi}_{{i}},\hat{\Pi}_{{j}})}{\bar{C} (\hat{\Pi}_{{i}},\hat{\Pi}_{{j}})},\\
    X_0 = 1 - X_1 &=& \bar{C} (\hat{\Pi}_{\text{0}},\hat{\Pi}_{\text{0}}) + \bar{C} (\hat{\Pi}_{\text{0}},\hat{\Pi}_{\text{1}}) \nonumber ,
\end{eqnarray}
with $i, j \in \{0, 1, +, -\}$ denote logical and Hadamard basis states~\cite{Rudnicki_2012}. Moreover, with the correlation matrix defined as \cite{Horst_2013}
\begin{equation}
    R_{{m,n}} = \bar{C}(\sigma_{{m}},           \sigma_{{n}}) - 4 C(\sigma_{{m}},           \sigma_{{n}}),
\end{equation}
where $\sigma_{{m}}$ and $\sigma_{{n}}$ are Pauli operators, one can calculate the entropic witness~\cite{Bovino2005,Bartkiewicz_2017} as 
\begin{equation}
    \mathrm{EW}(\hat{\varrho}) = \frac{1}{2}\Big[\text{Tr}(R) - 1 \Big]
\end{equation}
and also the CHSH witness in the form~\cite{MIRANOWICZ_2004,HORODECKI_1995}
\begin{equation}
    \mathrm{CHSH}(\hat{\varrho}) = \text{Tr}(R) - \text{min[eig }(R)] - 1.
\end{equation}

Purity of each state was also calculated. Note that purity can be directly measured using collective measurement on two copies of the investigated state using either the approach involving $n = 2$ nonlocal projections~\cite{Bovino2005,Bartkiewicz_2013}
\begin{eqnarray}
      P (\hat{\varrho}) = \text{Tr}[\hat{\varrho}_{\text{A}_{\text{1}}, \text{B}_{\text{1}}} \otimes \hat{\varrho}_{\text{A}_{\text{2}}, \text{B}_{\text{2}}} (1-2S)_{\text{A}_{\text{1}}, \text{A}_{\text{2}}} 
      \otimes (1-2S)_{\text{B}_{\text{1}}, \text{B}_{\text{2}}}] \nonumber
\end{eqnarray}
or in the entanglement swapping geometry following the idea to decompose nonlocal projection into a series of local measurements
\begin{equation}
    \hat{S}_{\text{A}_{\text{1}}, \text{A}_{\text{2}}} = \frac{1}{4}\Big(1- \sum_{{m}} \sigma_{{m}} \otimes           \sigma_{{m}}\Big)_{\text{A}_{\text{1}}, \text{A}_{\text{2}}}
\end{equation}
described in~\cite{Horst_2013}. 

\textit{Classification of states:} In the third step, each collective witness is paired with purity to obtain sets of feature vectors of length 2 to be then used for classification via machine learning (one set per witness). True labels for training were provided by the PPT criterion which is, for two-qubit states, a sufficient and necessary condition for entanglement. Note that in practical conditions the PPT criterion requires complete quantum state tomography to be performed and hence also the number of measurements exponentially growing with the size of the Hilbert space. Furthermore, negativity cannot be measured in a collective measurement on two copies of the investigated state~\cite{Bartkiewicz.PhysRevA.95.2017}.
Distribution of the feature vectors for all three witnesses colored based on their true label is visualized in Figures~\ref{fig_distr}~(a)--(c). Classification decision based solely on the analytical formulae of the three witnesses is visualized in these figures by dashed horizontal lines. Similarly, the states with purities $< 1/3$ that can not be entangled are delimited by a vertical dashed line~\cite{Rudnicki_2012}.

Incorporating purity into the classification of the states and simultaneously tuning the required sensitivity and selectivity is a task that can not be tackled by means of analytical calculations. We thus resort to a machine learning-based approach. For all three witnesses, the set of feature vectors was split into two halves, one used to train support vector machine classifiers (SVC) and the other to test their performance. After experimenting with the hyper-parameters, we have decided to use the radial basis function (RBF) kernels with $\gamma=1$. To speed-up the training, final classification of the test set is based on 11 hard-voting SVCs each trained on $1/11$ of the training instances. To observe complete ROC curves, SVC were trained with 12 different class-based penalties for misclassification of entangled states $w_\mathrm{e}$ ranging from $10^{0.75}$ to $10^{-1.15}$. Penalty for misclassification of separable states $w_\mathrm{s}$ was adjusted always so that $w_\mathrm{s} w_\mathrm{e} = 1$ to maintain regularization. Specific values of these penalties affect sensitivity and selectivity of the SVC. To benchmark sensitivity, we have calculated the APR (analytical positive rate), i.e. the percentage of entangled states that a given witness is capable of identifying solely based on the analytical threshold without the usage of purity. Sensitivity of each SVC is then characterized by its TPR (true positive rate) and the resulting improvement factor (IF)
\begin{equation}
\mathrm{IF} = \frac{\mathrm{TPR}}{\mathrm{APR}}.
\end{equation}
Selectivity, on the other hand, is directly described by the false positive rate (FPR). Note that decisions based solely on the analytical formulae of the witnesses (without the SVC) have $\mathrm{FPR}=0$. Complete programming code in Python 3 using the ThunderSVM library~\cite{wenthundersvm18} is available as Digital Supplement~\cite{digital_suppl}.

\section{Results}
For all combinations of the three witnesses and the 12 values of the class-based penalties $w_\mathrm{e}$, the SVC learned an optimal boundary between separable and entangled states in the feature space. Selection of these boundaries is depicted in Figures~\ref{fig_distr}~(a)--(c) together with the corresponding improvement factors IF. 
\begin{figure}
\includegraphics[scale=1]{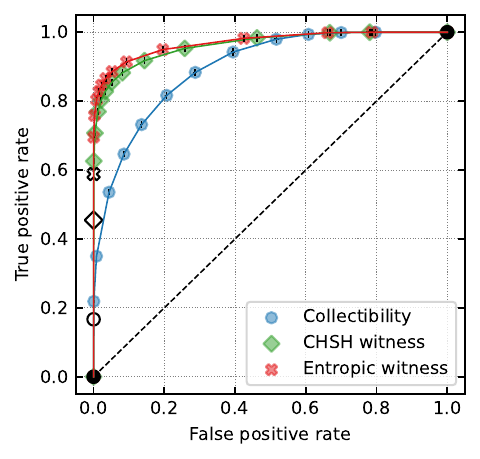}
\caption{\label{fig:roc}ROC curves for the purity-enhanced classification based on the three tested witnessed. Diagonal dashed line represents a naive (random) decision making while the filled markers correspond to SVCs trained with varying class-based penalties $w_\mathrm{e}$ and $w_\mathrm{s}$. Empty markers depict the APR, i.e. the TPR using solely the analytical formulae of the witnesses.}\label{fig_roc}
\end{figure}
Subsequently, we have obtained the confusion matrices for all these SVCs based on their decision on the test set (independent on the training set). TPR and FPR are then directly calculated from these matrices and plotted in Figure~\ref{fig_roc}. By splitting the test set into 50 batches and evaluating them independently, we were able to estimate standard deviation of all the presented quantities.
As a result we obtain complete ROC curves for the idea of purity-enhanced classifications. Areas under curve (AUC) for all three witnesses are numerically calculated to evaluate the performance of this method (see Table~\ref{tab_AUC_APR}). We have found out, that using this approach, one can improve the TPR typically by a factor of 1.3 by only sacrificing less than $1 \permil$ of selectivity (FPR < 0.001). Experimental imperfections and detection noise will inevitably cause even theoretically perfect witnesses to become imperfect. Considering that we believe that the price of a quite small selectivity drop is worth the tangibly increased sensitivity. 
Moreover, we have demonstrated that sensitivity can further be improved to near TPR close to 1 at the expense of decreased selectivity. This is well visible in the presented ROC curves whose AUC is, for all tested witnesses, greater than $0.9$. For all numerical results see Tables~\ref{tab_TPR_Col}--~\ref{tab_TPR_EW} in the Appendix. 
\begin{table} 
\caption{Values of areas under the curves (AUC) and analytical positive rates (APR) for tested witnesses. \label{tab_AUC_APR}}
\begin{ruledtabular}
\begin{tabular}{lcc}
                 & AUC & APR (\%) \\ \hline
Collectibility    & $0.902 \pm 0.002$ & $16.7 \pm 0.4$ \\
CHSH witness     & $0.965 \pm 0.001$ & $45.5 \pm 0.5$ \\
Entropic witness & $0.973 \pm 0.001$ & $58.8 \pm 0.6$ \\
\end{tabular}
\end{ruledtabular}
\end{table}

\section{Conclusions}
We have demonstrated that white noise significantly impairs the performance of collective entanglement witnesses. This effect was documented on the example of collectibility loosing the capability to detect entangled Werner states when noise level exceeds 13 \%. To mitigate this shortcoming, we have proposed to correlate the values of the collective witnesses with the state purity. Achieving this task analytically proves to be impractical, especially considering the need to tune detection selectivity. The solution lies in employing machine learning. Specifically we have implemented SVC models to optimize the decision boundary between entangled and separable states in the feature space spanned by the values of the entanglement witness and the states' purities. Our findings indicate that the range of detected entangled states can be expanded by a significant percentage at a minimal cost in sensitivity. Consequently, we believe that the idea of correlating entanglement witnesses with purity holds promise for practical near-future quantum communications, particularly in the context of entanglement swapping as showcased in this paper.

\section{Acknowledgements}
The authors thank Karol Bartkiewicz for useful discussion. The authors also acknowledge the CESNET for data management services.

\section*{Appendix}

\subsection{Preparing data sets \label{ap_prepare}} 

\begin{figure}[h]
\centering
\includegraphics[scale=1]{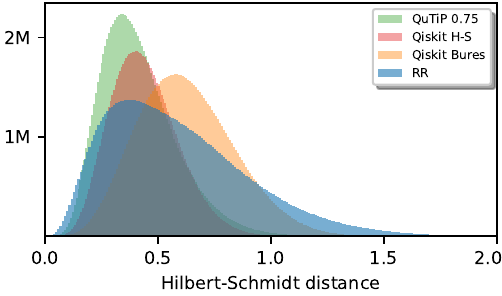}
\caption{Histogram of Hilbert-Schmidt distances of all combinations between 10.000 random states generated by  QuTiP library with parameter \texttt{density=0.75}, Qiskit library with two different settings of parameter \texttt{method='Hilbert-Schmidt','Bures'} and by random rotation (RR) method (\textit{200 bins}).}
\label{fig_HSdist}
\end{figure}

There are several methods to generate random density matrices. These methods usually employ some density matrix distance measure (Hilbert-Schmidt or Bures) and try to obtain uniform distribution of these distances over generated states, see Figure \ref{fig_HSdist}. Another method involves random diagonal matrices subjected to global unitary rotations~\cite{Maziero2015,Li2013}. There are dedicated libraries in Python implementing random state generation, for example \texttt{qiskit.quantum\_info.random\_density\_matrix()}~\cite{Qiskit} or \texttt{qutip.rand\_dm()}~\cite{JOHANSSON2013}.

These methods, in general, do not deliver states such as pure separable or pure entangled with sufficient prevalence because of their uncommon presence in the Hilbert space. The absence of pure states would overestimate the power of our method because we compare it with the analytical function of the witnesses that perform well on pure states.
To deal with this effect and make the conditions more challenging we decided to prepare a training data set with equally distributed purity where half of the states are entangled (negativity $N>0$) and the second not ($N=0$). This way we make sure that our method is fairly evaluated with respect to whatever conditions a user might have.

\begin{figure}[h]
\centering
\includegraphics[scale=1]{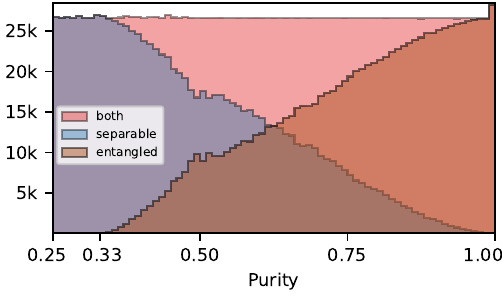}
\caption{Purity histogram of training and testing dataset consisting of 2 millions of states (\textit{75 bins}).}
\label{fig_Phist}
\end{figure}

\begin{figure}[h]
\centering
\includegraphics[scale=1]{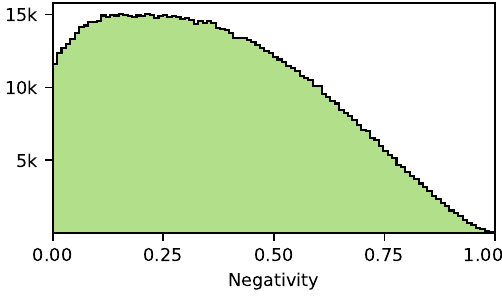}
\caption{Negativity histogram of training and testing dataset consisting of 1 million entangled states (\textit{100 bins}).}
\label{fig_Nhist}
\end{figure}

\begin{figure}[h]
\centering
\includegraphics[scale=1]{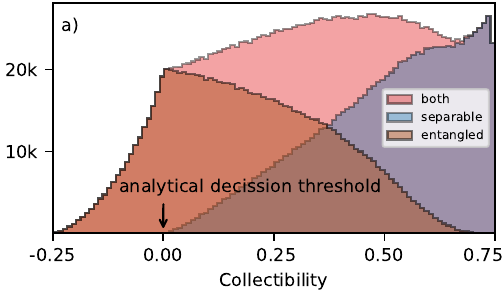}
\includegraphics[scale=1]{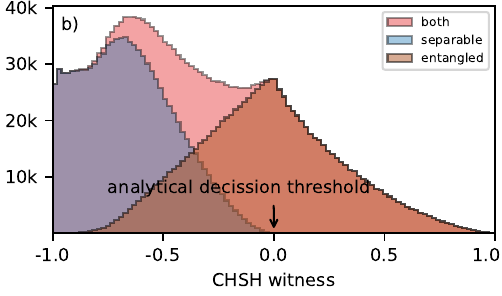}
\includegraphics[scale=1]{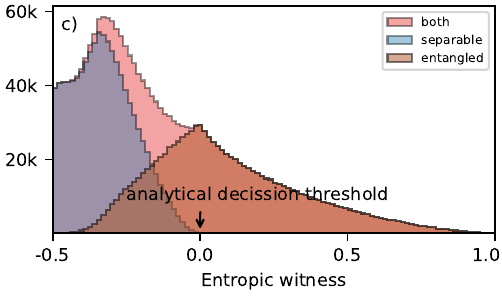}
\caption{Histograms of entanglement witnesses applied on training and testing dataset consisting of 2 millions of states, a) Collectibility, b) CHSH and c) Entropic witnesses (\textit{100 bins}).}
\label{fig_whist}
\end{figure}

For generation of states uniformly distributed in purity we used the method of random global  rotations because of its broad distribution in Hilbert-Schmidt distances. During states generation we control the values of purity (in 0.01 binning in interval [0.25,1]) and negativity (binary -- $N = 0$ or $N\neq 0$) and discard states which occur too frequently. Finally we obtain two million density matrices for which the values of negativity, purity, Collectibility, CHSH and Entropic witnesses were calculated. The histograms of these values are presented in Figures \ref{fig_Phist}, \ref{fig_Nhist} and \ref{fig_whist}.


\begin{figure}[h]
\centering
\includegraphics[scale=1]{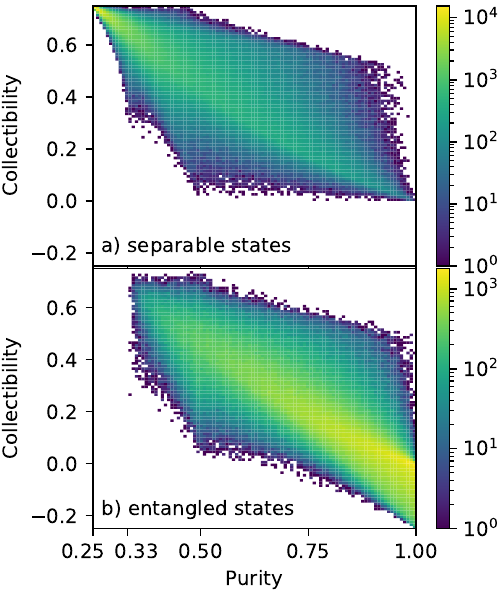}
\caption{Density maps of purity vs Collectibility of training and testing dataset consisting of a) 1 million separable and b) 1 million entangled states.}
\label{fig_2DCol}
\end{figure}

\begin{figure}[h]
\centering
\includegraphics[scale=1]{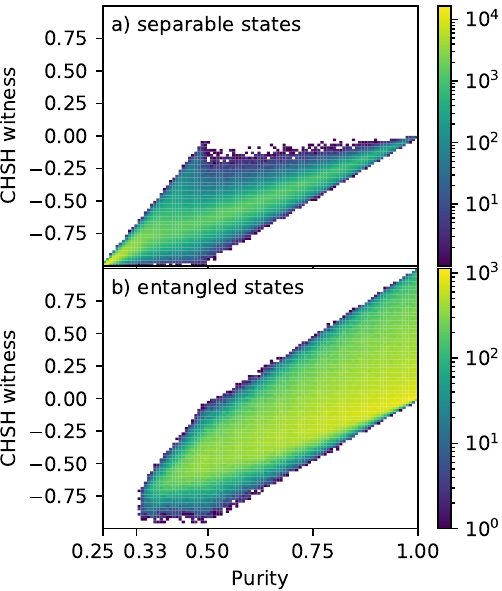}
\caption{Density maps of purity vs CHSH witness of training and testing dataset consisting of a) 1 million separable and b) 1 million entangled states.}
\label{fig_2DCHSH}
\end{figure}

\begin{figure}[h]
\centering
\includegraphics[scale=1]{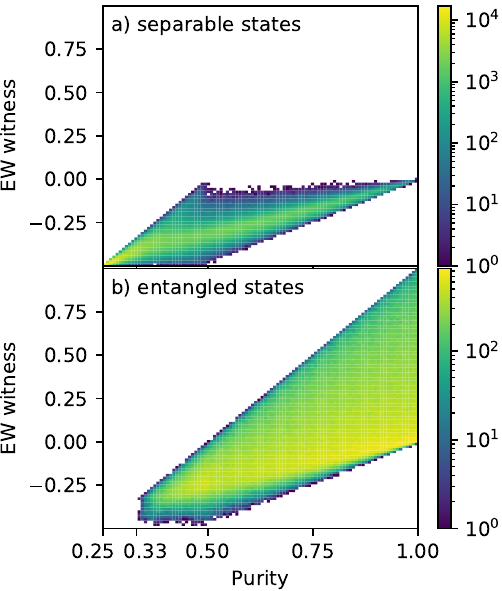}
\caption{Density maps of purity vs Entropic witness of training and testing dataset consisting of a) 1 million separable and b) 1 million entangled states.}
\label{fig_2DEW}
\end{figure}

Support vector machine (SVM) tries to find the border hyperplane between entangled and separable states knowing the values of purity and the analytical value of an entanglement witness. To assess the SVM decision we can visualize density maps (in logarithmic scale) of purity and witnesses under the consideration for separable and entangled states separately, see Figures \ref{fig_2DCol}, \ref{fig_2DCHSH} and \ref{fig_2DEW}.

\subsection{TPR and FPR values}

The decision of SVM results in a confusion matrix. On its diagonal lie true recognitions of entangled (true positive - TP) and separable (true negative - TN) states. Off-diagonal terms represent wrong assignment, false negative (FN) when entangled states were marked as separable and false positive (FP) whereas separable states were marked as entangled. True positive rate (TPR) and false positive rate (FPR) forming receiver operating characteristic (ROC) curve are calculated directly from the confusion matrix:

$$ \left( \begin{array}{cc} \mathrm{TP} & \mathrm{FN} \\ \mathrm{FP} & \mathrm{TN} 
\end{array} \right), \mathrm{TPR} = \frac{\mathrm{TP}}{\mathrm{TP}+\mathrm{FN}}, \mathrm{FPR} = \frac{\mathrm{FP}}{\mathrm{FP}+\mathrm{TN}}.$$

Upon training of the SVM the class-specific penalties $w_\mathrm{e}$ and $w_\mathrm{s}$ are tuned accordingly to homogeneously cover the entire interval of the ROC. For each point we can also calculate the improvement factor (IF) which quantifies how many more entangled states are recognized by the SVM when the value of purity is included in the decision. Naturally, this improvement is accompanied by the increments of misclassified separable states.

Results are summarized in Tabs. \ref{tab_TPR_Col} -- IV.
\begin{table}
\caption{Confusion matrices, values of true positive rates (TPR) and false positive rates (FPR) for different improvement factors (IF) of SVM learned on values of purity and the Collectibility. \label{tab_TPR_Col}}
\begin{ruledtabular}
\begin{tabular}{cccc}
IF & confusion matrix & TPR (\%) & FPR (\%) \\ \hline
$1.31\pm0.04$ & $\left( \begin{array}{cc} 109389  &  390616 \\ 342 & 499653 \end{array} \right)$ & $21.9\pm0.5$ & $0.07\pm0.02$ \\
$2.10\pm 0.06$ & $\left( \begin{array}{cc} 174947 & 325058 \\ 4120 & 495875 \end{array} \right)$ & $35.0\pm 0.6$ & $0.82\pm 0.08$ \\
$3.21\pm 0.09$ & $\left( \begin{array}{cc} 267851 & 232154 \\ 21849 & 478146 \end{array} \right)$ & $53.6\pm 0.9$ & $4.7\pm 0.2$ \\
$3.9\pm 0.1$ & $\left( \begin{array}{cc} 323142 & 176863 \\ 42900 & 457095 \end{array} \right)$ & $64.6\pm 0.9$ & $8.6\pm 0.3$ \\
$4.4\pm 0.1$ & $\left( \begin{array}{cc} 365933 & 134072 \\ 67859 & 432136 \end{array} \right)$ & $73.2\pm 0.9$ & $13.6\pm 0.3$ \\
$4.9\pm 0.1$ & $\left( \begin{array}{cc} 408305 & 91700 \\ 103367 & 396628 \end{array} \right)$ & $82.0\pm 1.0$ & $20.7\pm 0.4$ \\
$5.3\pm 0.1$ & $\left( \begin{array}{cc} 441720 & 58285 \\ 143770 & 356225 \end{array} \right)$ & $88\pm 1$ & $28.8\pm 0.5$ \\
$5.7\pm 0.1$ & $\left( \begin{array}{cc} 471245 & 28760 \\ 197114 & 302881 \end{array} \right)$ & $94\pm 1$ & $39.4\pm 0.6$ \\
$5.9\pm 0.1$ & $\left( \begin{array}{cc} 489877 & 10128 \\ 258399 & 241596 \end{array} \right)$ & $98\pm 1$ & $51.7\pm 0.8$ \\
$6.0\pm 0.1$ & $\left( \begin{array}{cc} 496859 & 3146 \\ 303811 & 196184 \end{array} \right)$ & $99\pm 1$ & $60.8\pm 0.9$ \\
$6.0\pm 0.2$ & $\left( \begin{array}{cc} 499555 & 450 \\ 350121 & 149874 \end{array} \right)$ & $100\pm 1$ & $70.0\pm 1.0$ \\
$6.0\pm 0.1$ & $\left( \begin{array}{cc} 500005 & 0 \\ 398691 & 101304 \end{array} \right)$ & $100\pm 1$ & $80\pm 1 $

\end{tabular}
\end{ruledtabular}
\end{table}

\begin{table}
\caption{Confusion matrices, values of true positive rates (TPR) and false positive rates (FPR) for different improvement factors (IF) of SVM learned on values of purity and the CHSH witness. \label{tab_TPR_CHSH}}
\begin{ruledtabular}
\begin{tabular}{cccc}
IF & confusion matrix & TPR (\%) & FPR (\%) \\ \hline
$1.38\pm 0.02$ & $\left( \begin{array}{cc} 313017  &  186988 \\ 302 & 499693 \end{array} \right)$ & $62.6\pm 0.8$ & $0.06\pm0.02$ \\
$1.56\pm 0.03$ & $\left( \begin{array}{cc} 353662 & 146343 \\ 1925 & 498070 \end{array} \right)$ & $70.7\pm 0.9$ & $0.39\pm 0.06$ \\
$1.69\pm 0.03$ & $\left( \begin{array}{cc} 384928 & 115077 \\ 6168 & 493827 \end{array} \right)$ & $77.0\pm 0.9$ & $1.2\pm 0.1$ \\
$1.77\pm 0.03$ & $\left( \begin{array}{cc} 401385 & 98620 \\ 10774 & 489221 \end{array} \right)$ & $80.3\pm 0.9$ & $2.2\pm 0.2$ \\
$1.82\pm 0.03$ & $\left( \begin{array}{cc} 414159 & 85846 \\ 16624 & 483371 \end{array} \right)$ & $83.0\pm 1.0$ & $3.3\pm 0.2$ \\
$1.88\pm 0.03$ & $\left( \begin{array}{cc} 427452 & 72553 \\ 25851 & 474144 \end{array} \right)$ & $85.0\pm 1.0$ & $5.2\pm 0.2$ \\
$1.94\pm 0.03$ & $\left( \begin{array}{cc} 441112 & 58893 \\ 40721 & 459274 \end{array} \right)$ & $88.0\pm 1.0$ & $8.1\pm 0.3$ \\
$2.02\pm 0.03$ & $\left( \begin{array}{cc} 458795 & 41210 \\ 71985 & 428010 \end{array} \right)$ & $92\pm 1$ & $14.4\pm 0.4$ \\
$2.10\pm 0.03$ & $\left( \begin{array}{cc} 476498 & 23507 \\ 128849 & 371146 \end{array} \right)$ & $95\pm 1$ & $25.8\pm 0.6$ \\
$2.16\pm 0.03$ & $\left( \begin{array}{cc} 491819 & 8186 \\ 231088 & 268907 \end{array} \right)$ & $98\pm 1$ & $46.2\pm 0.8$ \\
$2.20\pm 0.03$ & $\left( \begin{array}{cc} 499293 & 712 \\ 333931 & 166064 \end{array} \right)$ & $100\pm 1$ & $66.8\pm 1.0$ \\
$2.20\pm 0.03$ & $\left( \begin{array}{cc} 499978 & 27 \\ 390472 & 109523 \end{array} \right)$ & $100\pm 1$ & $78\pm 1 $
\end{tabular}
\end{ruledtabular}
\end{table}

\begin{table}
\caption{Confusion matrices, values of true positive rates (TPR) and false positive rates (FPR) for different improvement factors (IF) of SVM learned on values of purity and the Entropic witness. \label{tab_TPR_EW}}
\begin{ruledtabular}
\begin{tabular}{cccc}
IF & confusion matrix & TPR (\%) & FPR (\%) \\ \hline
$1.18\pm 0.02$ & $\left( \begin{array}{cc} 347708  &  152297 \\ 168 & 499827 \end{array} \right)$ & $69.5\pm 0.9$ & $0.03\pm0.02$ \\
$1.29\pm 0.02$ & $\left( \begin{array}{cc} 379263 & 120742 \\ 1227 & 498768 \end{array} \right)$ & $75.9\pm 0.9$ & $0.25\pm 0.05$ \\
$1.37\pm 0.02$ & $\left( \begin{array}{cc} 401933 &  98072 \\ 4285 & 495710 \end{array} \right)$ & $80.4\pm 0.9$ & $0.9\pm 0.1$ \\
$1.41\pm 0.02$ & $\left( \begin{array}{cc} 413866 & 86139 \\  7479 & 492516 \end{array} \right)$ & $82.8\pm 0.9$ & $1.5\pm 0.1$ \\
$1.44\pm 0.02$ & $\left( \begin{array}{cc} 423261 & 76744 \\ 11424 & 488571 \end{array} \right)$ & $84.7\pm 0.9$ & $2.3\pm 0.2$ \\
$1.47\pm 0.02$ & $\left( \begin{array}{cc} 432815 & 67190 \\ 17265 & 482730 \end{array} \right)$ & $86\pm 1$ & $3.5\pm 0.2$ \\
$1.51\pm 0.02$ & $\left( \begin{array}{cc} 442813 & 57192 \\ 26179 & 473816 \end{array} \right)$ & $88.6\pm 1.0$ & $5.2\pm 0.2$ \\
$1.55\pm 0.02$ & $\left( \begin{array}{cc} 457069 & 42936 \\ 46544 & 453451 \end{array} \right)$ & $91\pm 1$ & $9.3\pm 0.3$ \\
$1.61\pm 0.02$ & $\left( \begin{array}{cc} 474819 & 25186 \\  97820 & 402175 \end{array} \right)$ & $95\pm 1$ & $19.6\pm 0.5$ \\
$1.67\pm 0.02$ & $\left( \begin{array}{cc} 491159 & 8146 \\ 212344 & 287651 \end{array} \right)$ & $98\pm 1$ & $42.5\pm 0.8$ \\
$1.70\pm 0.02$ & $\left( \begin{array}{cc} 499314 & 691 \\ 330981 & 169014 \end{array} \right)$ & $100\pm 1$ & $66.2\pm 1.0$ \\
$1.70\pm 0.02$ & $\left( \begin{array}{cc} 499994 & 11 \\ 393410 & 106585 \end{array} \right)$ & $100\pm 1$ & $79\pm 1 $
\end{tabular}
\end{ruledtabular}
\end{table}
\FloatBarrier

\end{document}